\documentclass[aps,pra,twocolumn, showpacs]{revtex4}
\usepackage{graphicx}
\usepackage{dcolumn}
\usepackage{bm}

\begin{document}
\preprint{preprint}
\title{Optical Lattice in a High Finesse Ring Resonator}
\author{B. Nagorny}
\email{bnagorny@physnet.uni-hamburg.de}
\author{Th. Els\"{a}sser}
\author{H. Richter}
\author{A. Hemmerich}
\affiliation{Institut f\"{u}r Laser--Physik, Universit\"{a}t Hamburg, 
Jungiusstrasse 9, D--20355 Hamburg, Germany}

\author{D. Kruse}
\author{C. Zimmermann}
\author{Ph. Courteille }
\affiliation{Physikalisches Institut, Universit\"{a}t T\"{u}bingen, 
Auf der Morgenstelle 14, D--72076 T\"{u}bingen, Germany}

\date{\today}

\begin{abstract}

An optical lattice with rubidium atoms ($^{85}Rb$) is formed inside a ring resonator with a finesse of $1.8 \times 10^5$ and a large mode volume of 1.3~$mm^3$. We typically trap several times $10^6$ atoms at densities up to $10^{12} cm^{-3}$ and temperatures between 25 and 125~$\mu K$. Despite of the narrow bandwidth (17.3~kHz) of the cavity, heating due to intra--cavity intensity fluctuations is kept at a low level, such that the time evolution of the temperature is determined by evaporative cooling.

\end{abstract}


\maketitle

Arrays of cold atoms confined in the regularly spaced microscopic potentials of optical standing waves have become a model system of modern atomic physics. Such optical lattices have been extensively studied in the dissipative regime close to an atomic resonance \cite{greenberg} and in the far off-resonant case, loaded either with a magneto-optic trap \cite{grimm1}, or more recently, with Bose-condensed atomic samples \cite{kasevich}, \cite{haensch}. Possible applications range from atom lithography \cite{bell} to quantum information processing with neutral atoms \cite{brennen}, \cite{hemmerich}.

Intriguing novel aspects arise if the lattice is prepared inside an optical resonator with a finesse exceeding $10^5$ \cite{doherty}, \cite{horak}, \cite{hechenblaikner}. Such resonators provide a significant enhancement of the back--action of the atoms on the optical standing wave. Without resonator feedback, this back--action is tiny and can only be observed when the light is tuned close to an atomic resonance. In this case, the modification of the light field is well described by attributing a refractive index to the regularly spaced atoms \cite{weidemŸller}, \cite{raithel}. In presence of a resonator, however, multiple scattering has to be accounted for, yielding a strong coupling of the motion of atoms at distant lattice sites. While part of this coupling is conservative and may be employed in schemes for quantum gate operations, the finite lifetime of the photons in the resonator can lead to dissipation even in absence of spontaneous emission. As discussed in refs. \cite{hechenblaikner}, \cite{gangl} and \cite{vuletic}, this dissipation can be exploited for novel laser cooling schemes which apply to arbitrary polarizable particles, e.g. molecules. 

The motion of atoms inside high finesse resonators has been previously explored in the strong coupling regime in experiments aiming at large electric field strengths per photon obtained only for very small mode volumes below $10^{-4}$~$mm^3$ \cite{rempe}, \cite{kimble}. In this regime a few photons interact with a few atoms. In fact, the trapping of single atoms by single photons could be observed. In order to work with much larger sample sizes, the mode volume needs to be largely increased, yielding the complication of a correspondingly low resonator bandwidth. Previous experimental work on resonator based light traps with large mode volumes has explored a regime characterized by a finesse around a few hundred \cite{grimm2}.

In this article we report on the experimental realization of an optical lattice with $^{85}Rb$ atoms inside a ring resonator with a finesse of $1.8 \times 10^5$ and a large mode volume of 1.3~$mm^3$. For example, with 350~$\mu K$ deep potential wells we trap $4 \times 10^6$ atoms in this lattice at a peak density of $9 \times 10^{11} cm^{-3}$ and a temperature of  123~$\mu K$, which corresponds to a phase space density of $4.5 \times 10^{-6}$. For 100~$\mu K$ wells, $1.5 \times 10^6$ atoms are trapped at a peak density of $6.8 \times 10^{11} cm^{-3}$ and a temperature of  38~$\mu K$, i.e., the phase space density increases to $2 \times 10^{-5}$. A decrease of temperature with time accompanied by a decrease of the particle number is observed which can be well explained by a model based on evaporative cooling. Observations made, when ramping down the potential wells at different scanning speeds, support this interpretation. Heating due to intensity fluctuations is well controlled despite of the narrow bandwidth (17.3~kHz FWHM) of the cavity. From the spectral power density of the light transmitted through the cavity we calculate a time constant of 24~s for a temperature increase by a factor~$e$. Temperature measurements indicate even larger $e$-folding times above 100~s. Our intra--cavity optical lattice operates in the regime of strong collective interactions. This regime is characterized by $r N F \approx 1$, where $r$ is the field reflectivity per atom (related to the atomic polarizability by $\alpha = \varepsilon_0 r \lambda {w_0}^2$, with $w_0$~=~beam waist, $\lambda$~=~wavelength, and $\varepsilon_0$~=~dielectric constant), $N$ is the number of atoms in the cavity, and $F$ is the Finesse. In our present experiment cavity mediated cooling is not expected because the frequency of the incoupled light is kept exactly in resonance with the cavity.

\begin{figure}
\includegraphics[scale=0.3]{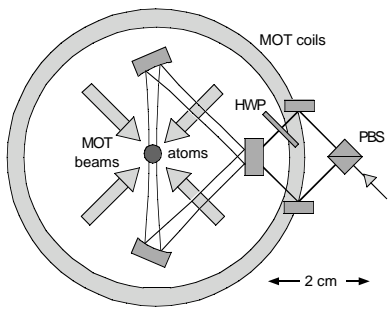}
\caption{ \label{Fig1}Sketch of the experimental setup. PBS=polarizing beam splitter, HWP=half wave plate. The entire unit is placed inside the vacuum chamber.}
\end{figure}

The experimental setup is sketched in Fig.1. We have chosen a ring geometry in view of future plans to explore the collective vibrational dynamics described in refs. \cite{hemmerich} and \cite{gangl}, which provides long decoherence times for vibrational modes with vanishing center of mass component. The triangular resonator is comprised of two curved high reflectors (0.8~ppm transmission, 3~ppm scattering loss, curvature radius = 200~mm) and a plane incoupling mirror (23~ppm transmission, 3~ppm scattering loss). It is placed such that the optical lattice is oriented vertically. The output beam of a grating stabilized laser diode, detuned to the red side of the rubidium D2 transition by 7.4~nm, is split by a polarizing beam splitter and coupled into both counterpropagating traveling wave modes. A cavity resonance linewidth of 17.3~kHz is measured by observing the exponential decay of the intra-cavity intensity ($\tau$~= 9.2~$\mu s$) after an abrupt termination of the incoupled light beam. From the 3.1~GHz free spectral range corresponding to the 97~mm round trip path length, a finesse of $1.8 \times 10^5$ is obtained. The sagittal and transversal $1/e^2$ mode diameters are 268~$\mu m$ and 258~$\mu m$ respectively. In order to stabilize the diode laser emission to the cavity resonance, we use a Pound-Drever--Hall technique with a servo bandwith of 3~MHz in the proportional feedback applied to the injection current \cite{schoof}. Both traveling modes have the same linear polarization perpendicular to the cavity plane and form an intensity grating. The triangular and thus perfectly planar geometry insures that no polarization rotation occurs during a round trip, which otherwise could degrade the finesse. The entire cavity setup, including the beam splitting unit, is placed inside the vacuum, in order to keep the optical path lengths between the polarizer cube and the incoupling mirror as short (and thus passively stable) as possible. With 60~$\mu W$ coupled into each traveling mode, we obtain a trap depth of 350~$\mu K$ at a spontaneous scattering rate of 40~$s^{-1}$. The corresponding axial and radial secular frequencies in the harmonic regime are 340~kHz and 460~Hz respectively. We can produce much deeper trap potentials. For example, 25 mW input power yields the Lamb-Dicke regime (vibrational frequency > recoil frequency) for the transverse and the strong confinement regime (vibrational frequency > natural linewidth) for the axial degree of freedom.
 
Loading of the lattice is accomplished with a magneto-optic trap (MOT) superimposed on the optical lattice which uses the $F=3 \rightarrow F=4$ cycling transition. The background pressure is in the low $10^{-10}$ mbar regime, however at  present our MOT lifetime is limited to 1.7~s by local contaminations emerging from the hot MOT coils placed at a few cm distance from the cold atoms inside the vacuum. After collecting $6 \times 10^8$ atoms the repumping beam (resonant with $F=2 \rightarrow F=3$) is shut off shortly before the MOT beams. This insures that initially all atoms captured by the optical lattice are pumped into the lower F=2 hyperfine level. We apply a ballistic expansion method in order to measure temperatures in the lattice. The lattice is suddenly turned off and, after a variable time, the sample is illuminated by a short (1~ms) light pulse slightly red detuned with respect to the resonance, and the fluorescence is recorded with a charge coupled device camera. From series of such expansion images at different expansion times we can  derive the spatial extension, the particle number, and the temperature of the initial sample. This methods works well for trapping times above 50~ms, for which MOT atoms, not captured by the lattice, have vanished and thus do not form an undesired background.   

\begin{figure}
\includegraphics[scale=0.45]{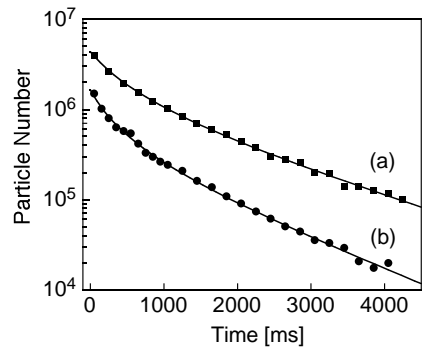}
\caption{ \label{Fig2} Time evolution of the number of trapped atoms. The potential well depth is 350~$\mu K$ in (a) and 100~$\mu K$ in (b). The solid lines show theoretical fits derived from eq.1}
\end{figure}

In Fig.2 we present measurements of the trap population versus time N(t)  for two different trap depths 350~$\mu K$ (a) and 100~$\mu K$ (b) respectively. The solid lines results from a model that accounts for density independent losses with a rate $\gamma$ and a density dependent loss term described by a parameter $\beta$ according to

\begin{equation}
\mbox{\.{N}} =  -\gamma \mbox{N} - \beta \int \rho^2(r)\,d^3r  ,
\end{equation}

where $\rho(r)$ is the density distribution of the atoms in the potential wells. The density independent loss rate $\gamma$ is comparable to that found in MOT decay experiments and should result from collisions with residual background gas. We attribute the difference in $\gamma$ in (a) and (b) to a slow gradual increase of the background pressure, when the apparatus is operated. Assuming a Gaussian distribution for the particle numbers across the potential wells and a thermal density distribution in the harmonic approximation inside each well, we find
$\gamma = 0.6~s^{-1}$ and $\beta  = 7.5 \times 10^{-12} s^{-1} cm^{3}$ for 350~$\mu K$  wells, and $\gamma = 0.76~s^{-1}$ and $\beta  = 1.7 \times 10^{-11} s^{-1} cm^3$ for 100~$\mu K$ wells. In Fig.3 we show the corresponding time evolution of the temperature. In both traces (a) and (b), instead of a temperature increase due to exponential heating by well depth fluctuations, a temperature decrease is observed which is faster during the first few hundred ms and subsequently continues at a lower rate. The solid lines result from a theoretical model based on evaporative cooling, which is discussed below.

\begin{figure}
\includegraphics[scale=0.45]{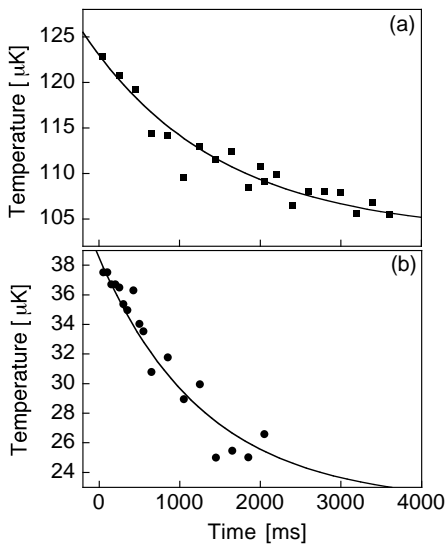}
\caption{ \label{Fig3} Time evolution of the temperature of trapped atoms. The potential well depth is 350~$\mu K$ in (a) and 100~$\mu K$ in (b). The solid lines show theoretical fits according to the model described in eq.3 with $\epsilon$ used as a fit parameter.}
\end{figure}

We can exclude hyperfine structure changing collisions (HSC) or photo associative collisions (PAC) as potential density dependent loss mechanisms by means of the following arguments. We expect that HSC loss only occurs for F=3 atoms. Although we initially prepare the atoms in the F=2 state, spontaneous scattering at a rate of 40~$s^{-1}$ populates the F=3 level within several ten ms. However, by illuminating the atoms with a weak depumping beam resonant with the $F=3 \rightarrow F=3$ transition, the time spent in the F=3 level can be kept well below the collision time, and thus no loss should occur. Experimentally, we do not observe a modification of the decay curves in Fig.2 in presence of the depumping beam. At several nm detuning PAC losses usually exhibit a sharp resonance behavior which facilitates a discrimination of such losses from other mechanisms. However, in view of the large intra--cavity power which could yield significant power broadening (see e.g. the spectra in ref. \cite{heinzen}) a brief consideration of such losses appears useful. The PAC loss rate $\Gamma_{PAC}$ is proportional to $\sigma_{PAC} \ \bar \rho \ v_{rms}$, where $\sigma_{PAC}$ is the PAC cross section, $\bar \rho$ is the mean particle density, and $v_{rms}$ is the root mean square velocity. The PAC cross section scales with the light intensity and thus the well depth $U_0$, the interaction time which is proportional to $v_{rms}^{-1}$, and the collision cross section $\sigma_{cc}$, i.e.
$\Gamma_{PAC} \propto U_0 \sigma_{cc} \bar \rho$.
Because in our experiment $U_0 = \eta \, k_B T$ with $\eta \approx 3$ independent of  $U_0$ or $T$, and in the harmonic approximation $ \bar \rho \propto N \eta^{\frac{3}{2}}$, we obtain $\Gamma_{PAC}\propto \eta^{\frac{5}{2}} N T \sigma_{cc}$.
The expression $T \sigma_{cc}$ is an increasing (near $T = 0$) or constant (near the unitarity limit) function of T and thus also of $U_0$.
Since the particle number N decreases with decreasing $U_0$ in our experiment, we expect a decrease of PAC losses for a decrease of the well depth in contrast to our observations in Fig.2.

The observed temperature decrease and particle loss is consistent with an explanation based on evaporative cooling. We adopt a simple model for evaporation based on the principle of detailed balance \cite{ketterle} which predicts a particle loss rate $\Gamma_{ev} = \bar \rho \, \sigma_{esc} \, v_{rms} \, \eta e^{-\eta}$, where $\sigma_{esc}$ denotes the elastic scattering cross section, and $\bar \rho$ is the mean particle density in the lattice. This model assumes that after initial preparation at temperature T inside a harmonic potential, all atoms with an energy larger than $U_0$ are allowed to escape. If the temperature decrease with time is neglected (i.e. $v_{rms} \eta e^{-\eta}$ is constant in time), in analogy to eq.1, we obtain a corresponding value $\beta_{esc}$ = $\sigma_{esc} \, v_{rms} \, \eta e^{-\eta}$. In our experiment we need to consider collisions between atoms populating all Zeeman components of both hyperfine ground states with most of the possible collision processes involving singlet and triplet contributions. The large scattering length values found for $^{85}Rb, 5S_{1/2}, F=3$ \cite{roberts} lead us to employ an effective scattering cross section $\sigma_{esc}$ approximated by the unitarity limit $4 \pi \hbar^2 / \mu^2 \delta v_{rms}^2$, where $\mu$ = m/2 is the reduced mass and $\delta v_{rms}$ is the root mean square relative velocity. This yields 

\begin{equation}
 \beta_{esc} = \frac{8 \pi \hbar^2 \eta^{\frac{3}{2}} e^{-\eta}}{ \sqrt{3 U_0 m^3} }.
\end{equation}

Since in our experiment the value of $\eta$ does not depend on the well depth $U_0$, the above expression predicts an increase of $\beta_{esc}$, if the well depth is decreased, in accordance with the observations in Fig.2. Inserting the initial temperatures 123~$\mu K$ and 38.5~$\mu K$ and the corresponding well depths 350~$\mu K$ and 100~$\mu K$, we can evaluate $\beta_{esc}$ to be $1.2 \times 10^{-11} s^{-1} cm^3$ and $2.3 \times 10^{-11} s^{-1} cm^3$ respectively. These values are in reasonable agreement with those obtained above from the observations in Fig.2. The corresponding scattering cross sections agree within 10 \% with those calculated for $5S_{1/2}, F=3$ in ref. \cite{burke}.

We may use the decay curves in order to model the temporal evolution of the temperatures observed in Fig.3. The total kinetic energy at time t is given by
$N(t) W(t) = N(0) W(0) - N_1(t) W(0) - N_2(t) \bar W$, where
$N(t) = N(0) \ exp(-\gamma t) / (1+ \xi \, [1 - exp(-\gamma t)])$ is the atom number at time $t$ (obtained by solving eq.1), $N_1(t)=N(0) \ (1 - exp(-\gamma t))$ is the number of atoms lost due to collisions with hot background gas particles, $N_2(t)=N(0) - N_1(t) - N(t)$ is the number of particles lost due to elastic two-body collisions, $W(t)$ is the mean kinetic energy per particle, and $\bar W$ is the mean kinetic energy per particle removed by evaporation. The two-body loss parameter $\xi$, related to the value of $\beta$ in eq.1 and the peak density $\rho_{peak}$ by $\xi$=$\beta \rho_{peak}/4 \gamma$, is taken from the theoretical fits to the data of Fig.2. Since $\eta$ is sufficiently larger than one, we neglect the kinetic energy of particles after evaporation and assume, that the total energy loss per evaporated particle is $U_0$. Thus, the mean kinetic energy per particle removed by evaporation is roughly approximated by 
$\bar W = U_0 -  \bar U_{\eta}$ where $\bar U_{\eta}$ is the mean potential energy at the initial temperature T(0) within the trapping volume given by $U(x,y,z) < U_0$. Using $W= \frac{3}{2} k_B T$, we obtain 

\begin{eqnarray}
T(t) = T(0) \, \left( 1 - \epsilon \, \xi \left( 1 - e^{- \gamma t} \right)  \right), {}\nonumber \\
\epsilon = 
\frac{2}{3} \eta -1 - \frac{8}{3\sqrt{\pi}} \int_{0}^{\sqrt{\eta}} r^4 e^{-r^2} dr.
\end{eqnarray}

For trace (a) of Fig.3 we have $\xi=2.80$ and $\eta=2.85$ and thus $\epsilon=0.23$. Similarly, for trace (b), $\xi=3.72$, $\eta=2.63$, and $\epsilon=0.14$. The solid lines in Fig.3 are obtained by using $\epsilon$ as a fit parameter yielding $\epsilon=0.057$ for (a) and $\epsilon=0.12$ for (b). Note that the fitted values of $\epsilon$ deviate from the calculated ones only by a factor 4 for trace (a) and a factor 1.2 for trace (b) despite of the simplicity of our model. 

We can enhance evaporative cooling by slowly ramping down the potential well depth. For example, when we lower the potential from 350~$\mu K$ to 147~$\mu K$ in 70~ms, the temperature decreases to 64~$\mu K$. A 10 ms ramp, which is too fast for rethermalization, merely yields 81~$\mu K$ as expected from adiabatic cooling. Recall, that for an adiabatic change of the well depth from an initial value $U_i$ to a final value $U_f$ in the harmonic approximation $T_f= T_I \, \sqrt{U_f / U_i}$, which yields 79.7~$\mu K$, if specified for the situation considered here.

The observations of Fig.3 in connection with eq.3 let us determine an upper bound for the heating due to intensity fluctuations inside the resonator. This heating can be described by a simple mechanical model based on parametric excitation \cite{savard} which predicts a temporal increase of the mean kinetic energy according to $\dot W$ =  $\gamma_a W_a + \gamma_r W_r$, where $\gamma_a$ and $\gamma_b$ are the axial and radial  heating rates, and $W_a$ and $W_r$ are the axial and radial mean kinetic energies.
Assuming thermal equilibrium, i.e. $W_r / 2$ = $W_a$ = $W/3$, yields 
$\dot W$ =  $\gamma_{tot} W$ with the total heating rate $\gamma_{tot}=\frac{1}{3} \, (\gamma_{a}+ 2 \gamma_{r})$. We may incorporate this heating mechanism in eq.3, writing 
$\dot T(t) = - \epsilon \ \xi \ \gamma \ exp(- \gamma t) \ T(0) + \gamma_{tot} T(t)$. 
Since we observe a negative slope for T(t) on the entire time axis in trace (a) of Fig.3, we may conclude $0 > - \epsilon \ \xi \ \gamma \ exp(-\gamma t) T(0) + \gamma_{tot} T(t)$ for 
$0 \leq t  \leq 4 s$. Evaluating this relation for t = 4 s by means of Fig.3, we obtain 
$\gamma_{tot} < 0.01 \, s^{-1}$. 
Alternatively, $\gamma_a$ and  $\gamma_r$ can be calculated from the spectral power density of the light transmitted through the cavity at twice the axial and radial harmonic frequencies. Assuming thermal equilibrium yields $\gamma_{tot}= 0.041 \, s^{-1}$, which is a factor 4 above the upper limit discussed above. This discrepancy, however, is not surprising, because in Fig.3 radial temperatures are shown, while the heating mainly acts on the axial degree of freedom ($\gamma_a \gg \gamma_r$). As time evolves in Fig.3 the thermalization time increases and the axial and radial temperatures begin to deviate.

In summary, we have discussed the experimental realization of an optical lattice with rubidium atoms inside a ring-cavity with a finesse of 1.8~$\times 10^5$ and we have characterized the relevant trapping parameters. We have shown that well depth fluctuations can be kept at a low level such that the time evolution of the temperature is determined by evaporative cooling. Our cavity operates under conditions, where collective vibrations involving distant atoms should become visible, thus opening up a new exciting regime of atom--cavity dynamics.  

.
\begin{acknowledgments}
This work has been supported in part by the Deutsche 
Forschungsgemeinschaft under contract number He2334/2-3. We are grateful for discussions with Helmut Ritsch.
\end{acknowledgments}

\end{document}